# A quantum method to test the existence of consciousness

Rui Qi

Institute of Electronics, Chinese Academy of Sciences

17 Zhongguancun Rd., Beijing, China

E-mail: rg@mail.ie.ac.cn

## Introduction

As we know, "Who can be said to be a conscious being?" is one of the hard problems in present science, and no method has been found to strictly differentiate the conscious being from the being without consciousness or usual matter. In this paper, we will present a strict physical method based on revised quantum dynamics to test who can be said to be a conscious being, and the principle is to use the distinguishability of nonorthogonal single states.

## Revised quantum dynamics

As to the evolution of the wave function during quantum measurement, present quantum theory provides by no means a complete description. The projection postulate is just a makeshift, while the concrete dynamical process of the projection is undoubtedly one of the most important unsettled problems in quantum theory. Recently the resulting revised quantum dynamics ( Ghiradi et al, 1986; Pearle, 1989; Diosi, 1989; Ghiradi et al, 1990; Penrose, 1996; Gao, 1999a; Gao, 2000b; Gao, 2001b ) are deeply studied, in which the linear evolution equation of the wave function is replaced by stochastic linear or nonlinear equation. Presently, even if the last theory has not been found, but one thing is certain for the revised quantum dynamics, i.e. the collapse process is one kind of dynamical process, and it will take a finite time interval to finish. Our method in this paper only relies on this common character of revised quantum dynamics.

## The method to test the existence of consciousness

Now, we will demonstrate how to test the existence of consciousness in the framework of revised quantum dynamics. The concrete method is to use the distinguishability of nonorthogonal single states ( Gao, 1999b; Gao, 2000a; Gao, 2000b; Gao, 2001a).

As we know, the usual measurement using physical measuring device can't distinguish the nonorthogonal single states in revised quantum dynamics, as well as in present quantum theory. But, if the physical measuring device is replaced by a conscious being, we will demonstrate that it may distinguish the nonorthogonal single states in the framework of revised quantum dynamics. Thus the existence of consciousness can be tested by use of this physical method.

We assume the states to be distinguished are the following nonorthogonal single states $\psi_1$ and $\psi_1 + \psi_2$, and the initial perception state of the conscious being is $c_0$. Then after interaction the corresponding entangled state of the whole system is respectively $\psi_1 c_1$ and $\psi_1 c_1 + \psi_2 c_2$, where $c_1$ and $c_2$ is respectively the perception state of the observer for the states $\psi_1$ and $\psi_2$. We assume the observer satisfies the QSC condition (Gao Shan, 1999b; Gao Shan, 2000a), i.e. the perception time of the observer for the definite state $\psi_1 c_1$, which is denoted by $t_P$, is shorter than the dynamical collapse time for the superposition state $\psi_1 c_1 + \psi_2 c_2$, which is denoted by $t_C$ [1], and the time difference $\Delta t = t_C - t_P$ is large enough for the observer to identify. Then the observer can perceive the measured state $\psi_1$ or his own state $c_1$ after time interval $t_P$, while for the measured superposition state $\psi_1 + \psi_2$, only after the time interval $t_C$ can the observer perceive the collapse state $\psi_1$ or $\psi_2$, or his own corresponding state $c_1$ or $c_2$. Since the observer can also be conscious of the time difference between $t_P$ and $t_C$, he can easily distinguish the measured nonorthogonal single states $\psi_1$ and $\psi_1 + \psi_2$. Thus the distinguishability of the nonorthogonal single states can be used as a quantum method to differentiate man and machine, or to test the existence of consciousness.

## Further discussions

In order to understand the unusual conclusion, we will further analyze the above demonstrations. As we know, it is still unclear that what the perception of the observer in the entangled state $\psi_1 c_1 + \psi_2 c_2$ is. Albert had analyzed the similar situation in detail (Albert, 1999b). He called such quantum observer John. He concluded that John's perception is not the same as $c_1$ and $c_2$, and denoted that the perception may be very strange. In the following, we will further demonstrate that the above conclusion is irrelevant to the concrete perception of the observer in the superposed state.

First, we assume that only after the collapse the definite perception about the input superposition state can appear, which is a well-accepted fact in quantum mechanics Since the observer can be aware of his perception instant, he can also be aware of the collapse instant. Then when the observer satisfies the above assumed QSC condition, the awareness of collapse instant will permit him to distinguish the input states $\psi_1+\psi_2$ and $\psi_1$.

---

[1] It should be noted that, since the collapse time of a single superposition state is an essentially stochastic variable, which average value is $t_c$, we should consider the stochastic distribution of the collapse time in a strict sense, i.e. a small number of single states is needed for practical application. In the following discussions, we always simply take the collapse time as the average value $t_c$ unless state otherwise.

Secondly, we assume that the above well-accepted fact is not true, i.e. the observer can have some definite perception about the input superposition state before the collapse happens. Now we will demonstrate that the observer can also be aware of the collapse instant for this situation, thus the observer can also distinguish the input states $\psi_1+\psi_2$ and $\psi_1$ when satisfying the QSC condition.

(1). If the definite perception of the observer in the superposed state $\psi_1 c_1+\psi_2 c_2$ is neither $c_1$ nor $c_2$, then the observer can be aware of the collapse instant, since after the collapse instant the perception turns to be $c_1$ or $c_2$, which is different from that before the collapse instant, and the observer can be aware of the change of his perception.

(2). If the definite perception of the observer in the superposed state $\psi_1 c_1+\psi_2 c_2$ is $c_1$, then due to the randomness of the collapse result, the observer can still be aware of the collapse instant for one half of the situations, since after the collapse instant the perception will turn to be $c_2$ with probability 1/2.

(3). If the definite perception of the observer in the superposed state $\psi_1 c_1+\psi_2 c_2$ is $c_2$, the demonstration is the same as that of (2).

(4). If the definite perception of the observer in the superposed state $\psi_1 c_1+\psi_2 c_2$ is random[2], i.e. one time is $c_1$, another time is $c_2$, then due to the independent randomness of the collapse process, the observer can still be aware of the collapse instant with non-zero probability, since the perception after the collapse instant will be different from that before the collapse instant with non-zero probability.

Thus we have demonstrated that if only the observer satisfies the QSC condition, he can distinguish the measured nonorthogonal single states. The conclusion is irrelevant to the concrete perception of the observer in the superposed state.

## The rationality of QSC condition

Lastly, we will demonstrate that the QSC condition is not irrational, and can be satisfied in essence, i.e. there should exist some kind of conscious beings satisfying the condition in Nature.

First, the perception time of the conscious being is mainly determined by the structure of his perception part, while the dynamical collapse time of the observed superposition state during perception is mainly determined by the energy involved for perception. It is evident that the structure and energy for perception can't determine each other uniquely, or we can say, they are relatively independent. Thus the corresponding perception time and dynamical collapse time are also relatively independent. Then it is natural for some kind of conscious beings the above QSC condition is satisfied, and for other conscious beings the above QSC condition is not satisfied.

---

[2] This presumption may be extremely impossible.

Secondly, with the natural selection the structure of the perception part of the conscious being will turn more and more complex, and the perception time will turn shorter and shorter. On the other hand, the energy involved for perception will turn less and less, and the dynamical collapse time will turn longer and longer. Then there will appear more conscious beings satisfying the QSC condition with the natural evolution[3].

In one word, it is reasonable that QSC condition is satisfied by some kind of conscious beings, i.e. for some kind of conscious beings the perception time for the definite state $\psi_1$ is shorter than the perception time or dynamical collapse time of the perceived superposition state $\psi_1 + \psi_2$, and the time difference is large enough for the conscious beings to identify. Thus even if our human being can not satisfy this condition, other conscious beings may satisfy this condition. In fact, some evidences have indicated that our human being can satisfy this condition (Duane et al, 1965; Grinberg-Zylberbaum et al, 1994), for example, the subjects can hold the superposition state for a long time, say at least several minutes, in the experiments performed by Grinberg-Zylberbaum et al (Grinberg-Zylberbaum et al, 1994). This denotes that the collapse time of the superposition state, which is the same as the holding time of the superposition state, is much longer than the perception time, which is generally in the level of milliseconds.

## Conclusions

We show that the conscious being may distinguish the nonorthogonal single states when satisfying the QSC conditions, while the physical measuring device can't. This indicates that the distinguishability of nonorthogonal single states can be used to test the existence of consciousness.

---

[3] Owing to the availability of superluminal communication, satisfying the QSC condition will be undoubtedly helpful for the existence and evolution of the conscious beings.